\RequirePackage{lineno} 
\documentclass[reprint,twocolumn,aps,prc,showpacs,superscriptaddress,groupedaddress,nofootinbib]{revtex4}
\DeclareMathAlphabet{\mathpzc}{OT1}{pzc}{m}{it}
\usepackage{lineno}

\usepackage{graphicx}
\usepackage{amssymb}
\usepackage{amsmath}
\usepackage{amsthm}
\usepackage[percent]{overpic}
\usepackage{float}
\usepackage{array}
\usepackage{multirow}
\usepackage{epsfig}
\usepackage{color}
\usepackage{verbatim}

\begin{document}

\title{Experimental neutron capture data of $^{58}$Ni from the CERN n\_TOF facility}

\author{P.~\v{Z}ugec}\thanks{Corresponding author: pzugec@phy.hr}\affiliation{Department of Physics, Faculty of Science, University of Zagreb, Croatia}%
\author{M.~Barbagallo}\affiliation{Istituto Nazionale di Fisica Nucleare, Bari, Italy}%
\author{N.~Colonna}\affiliation{Istituto Nazionale di Fisica Nucleare, Bari, Italy}%
\author{D.~Bosnar}\affiliation{Department of Physics, Faculty of Science, University of Zagreb, Croatia}%
\author{S.~Altstadt}\affiliation{Johann-Wolfgang-Goethe Universit\"{a}t, Frankfurt, Germany}%
\author{J.~Andrzejewski}\affiliation{Uniwersytet \L\'{o}dzki, Lodz, Poland}%
\author{L.~Audouin}\affiliation{Centre National de la Recherche Scientifique/IN2P3 - IPN, Orsay, France}%
\author{V.~B\'{e}cares}\affiliation{Centro de Investigaciones Energeticas Medioambientales y Tecnol\'{o}gicas (CIEMAT), Madrid, Spain}%
\author{F.~Be\v{c}v\'{a}\v{r}}\affiliation{Charles University, Prague, Czech Republic}%
\author{F.~Belloni}\affiliation{Commissariat \`{a} l'\'{E}nergie Atomique (CEA) Saclay - Irfu, Gif-sur-Yvette, France}%
% ...
%

\collaboration{The n\_TOF Collaboration (www.cern.ch/ntof)}  \noaffiliation

\begin{abstract}
The $^{58}$Ni($n,\gamma$) cross section has been measured at the neutron time of flight facility n\_TOF at CERN, in the energy range from 27 meV up to 400 keV. In total, 51 resonances have been analyzed up to 122 keV. Maxwellian averaged cross sections (MACS) have been calculated for stellar temperatures of $kT~=~5~-~100$~keV with uncertainties of less than 6\%, showing fair agreement with recent experimental and evaluated data up to $kT$ = 50 keV. The MACS extracted in the present work at 30 keV is $34.2\pm0.6_\mathrm{stat}\pm1.8_\mathrm{sys}$ mb, in agreement with latest results and evaluations, but 12\% lower relative to the recent KADoNIS compilation of astrophysical cross sections. When included in models of the $s$-process nucleosynthesis in massive stars, this change results in a 60\% increase of the abundance of $^{58}$Ni, with a negligible propagation on heavier isotopes. The reason is that using both, the old or the new MACS, $^{58}$Ni is efficiently depleted by neutron captures.
\end{abstract}

\pacs{25.40.Lw, 25.40.Ny, 27.40.+z}
\maketitle

\section{Introduction}

The ($n, \gamma$) cross section of $^{58}$Ni is of interest for applications in nuclear technologies and astrophysics. In nuclear technology $^{58}$Ni is an important constituent of structural materials. Under neutron exposure, it contributes to the long-term radiation hazard through the production of $^{59}$Ni with a half-life of $7.5\times10^4$ years.

In astrophysics the ($n, \gamma$) cross section of $^{58}$Ni is required to characterize the role of $^{58}$Ni in the reaction network of the slow neutron capture process ($s$-process), which is responsible for about half of the abundances of the elements heavier than iron \cite{bibWW}. The $s$-process operates during the He and C burning stages of stellar evolution at temperatures between 0.1 and 1 GK (equivalent to thermal energies from $kT=8$ to 90 keV). For the determination of the effective neutron capture rate in a stellar environment, the energy differential ($n, \gamma$) cross section $\sigma(E_n)$ has to be known over a wide energy range up to several hundred keV. Folding the differential cross section with the thermal Maxwell-Boltzmann spectrum yields the Maxwellian averaged cross section (MACS) characteristic of the stellar environment. 

With an isotopic abundance of 68\%, $^{58}$Ni is among the prominent members of the Fe abundance peak and represents a secondary seed isotope for nucleosynthesis in the $s$-process. The solar Ni abundance has been produced along with the Fe-peak elements by core-collapse supernovae of massive stars \cite{bibZ} as well as by thermonuclear supernovae of type Ia \cite{bibAA} in complete nuclear statistical equilibrium during $\alpha$-rich freeze-out. According to spectroscopic observations at different metallicities, about 70\% of the solar Ni is made by Ia supernovae and 30\% by massive stars \cite{bibBB}.

Several measurements of the $^{58}$Ni ($n, \gamma$) cross section are available in EXFOR \cite{bibO}, but only few extend over a wide astrophysically relevant neutron energy range. Recently, a new measurement was performed from 100~eV to 600 keV by Guber \emph{et al.} at ORELA \cite{bibN}, indicating that the MACS data based on the evaluated cross sections from major libraries -- including ENDF/B-VII.0 \cite{bibSS} -- were overestimated by as much as 20\%. This difference may be attributed to the fact that -- among several other sources of experimental data -- ENDF/B-VII.0 relies on the capture measurement by Perey \emph{et al.} \cite{bibZZ} who have used C$_6$F$_6$ detectors, exhibiting higher neutron sensitivity than C$_6$D$_6$ detectors, which have been used by Guber \emph{et al}. A global decrease in the capture cross section -- compared to the past evaluations -- is also supported by the recent activation measurement by Rugel \emph{et al.} \cite{bibAAA}. Based on these results, a new evaluation was prepared for ENDF/B-VII.1 \cite{bibTT}, including also the high resolution transmission measurement by Brusegan \emph{et al.} \cite{bibM} and the new thermal value of Raman \emph{et al.} \cite{bibMM}. The present work was motivated by the sizable differences between the experimental cross section data of $^{58}$Ni \cite{bibO}, aiming at a high-resolution measurement from thermal to about 400 keV neutron energy.

\section{Experimental setup}

The radiative neutron capture cross section of $^{58}$Ni was measured at the neutron time of flight facility n\_TOF at CERN. At n\_TOF -- one of the most luminous neutron sources presently available -- neutrons are produced by an intense pulsed beam of 20 GeV/$c$ protons impinging on a massive cylindrical Pb target. Typical beam conditions are a repetition rate in multiples of 1.2 s and a proton pulse width of 7 ns. On average, 300 neutrons are produced per proton, resulting in a total of $2\times10^{15}$ neutrons per pulse. Besides 1 cm layer of water used for cooling the Pb target, 4 cm layer of borated water is used for moderating the initially fast neutrons. The final neutron spectrum spans the energy range from thermal to several GeV. The production of 2.2 MeV $\gamma$-rays from radiative neutron capture on hydrogen is also minimized by using borated water as the main moderator. Although strongly suppressed by the borated water, the thermal neutron component still ensures  a significant capture yield.

An evacuated beam line is connected to the spallation target and moderator. Neutrons, charged particles and intense $\gamma$-radiation outside the beam line are attenuated by the massive concrete walls and 3.5 m thick iron shielding. The charged particles from inside the beam line are removed by a 1.5 T sweeping magnet. The neutron beam is shaped by collimators at 137 and 178 m from the spallation target. The aperture of the second collimator was 19 mm, corresponding to the capture mode of the n\_TOF facility. Behind the experimental area, at about 185 m, the beam line continues for another 10 m to the beam dump, in order to minimize the back-scattering of neutrons. Further details on the facility may be found in Refs. \cite{bibB,bibC}.

The properties and dimensions of the metallic $^{58}$Ni sample, supplied by Chemotrade, are listed in Table \ref{tab1}. Due to high isotopic enrichment of 99.5\%, the sample contained only 0.48\% of $^{60}$Ni, 0.01\% of $^{61}$Ni and 0.005\% of both $^{62}$Ni and $^{64}$Ni.

\begin{table}[b!]
\caption{Characteristics of the $^{58}$Ni sample.}
\label{tab1}
\begin{tabular}{c>{\quad}c>{\;}cc} 
\hline\hline
Mass&Diameter&Thickness&Enrichment\\
\hline
2.069 g&19.91 mm&0.72 mm&$99.5\%$\\
&&($6.9\times 10^{-3}$ at/barn)&\\
\hline\hline
\end{tabular}
\end{table}

The prompt capture $\gamma$-rays were detected  by two optimized C$_6$D$_6$ liquid scintillation detectors -- a commercial detector (Bicron) that had been modified in order to conform with the specific requirements of the neutron capture experiment \cite{bibD}, and a custom built version (FZK detector). The scintillation liquid was contained within the cylindrical compartment of 618~ml for Bicron and 1027~ml in case of FZK. These volumes were low enough to provide sufficiently low $\gamma$-ray detection efficiency, required for detecting no more than one $\gamma$-ray from $\gamma$-ray cascades emitted in the neutron capture reactions. The detectors were mounted 8.2 cm upstream of the sample to reduce the effect of in-beam $\gamma$-rays, and at a distance of 6.8 cm from the center of the neutron beam line.

C$_6$D$_6$ detectors are characterized by exceptionally low neutron sensitivity. The use of carbon fiber for the evacuated beam line and for the supports further minimized the probability for neutron capture by the experimental apparatus itself. In this manner the neutron sensitivity of the experimental setup -- which can be defined as the ratio $\varepsilon_n/\varepsilon_\gamma$ between the efficiency $\varepsilon_n$ for detecting scattered neutrons (through secondary $\gamma$-rays produced by their interaction in the experimental area) and the efficiency $\varepsilon_\gamma$ for detecting capture events -- was kept as low as manageable.

The neutron flux was monitored during the measurement by means of SiMon -- a silicon-based neutron beam monitor \cite{bibRR} based on the $^6$Li($n,t$)$\alpha$ reaction. For determination of the capture yield, the evaluated neutron flux $\Phi(E_n)$ described in Ref. \cite{bibL} was used. With the second collimator in capture mode, the flux of neutrons below 400 keV, integrated over the nearly Gaussian beam profile at the sample position amounts to $2.2\times10^5$ neutrons per pulse.

The electronic signals were recorded by a high-performance digital data acquisition system, based on 8-bit flash-ADC units operating at a sampling rate of 500 MHz. A memory buffer of 48 MB allows us to measure up to 96 ms  in neutron time-of-flight, corresponding to a  neutron energy of 20 meV. However, data analyses are limited to the region above 27 meV, where the neutron flux has been reliably evaluated. The digitized signals were processed offline by specially developed and optimized data analysis algorithms. A detailed description of the digital data acquisition system may be found in Ref. \cite{bibE}.

\section{Calculation of the yield}

The aim of the data analysis is to determine the neutron capture yield of the $^{58}$Ni($n$,$\gamma$)$^{59}$Ni reaction, which is essential for extracting the pointwise cross section and the resonance parameters.

The energy calibration of the C$_6$D$_6$ detectors was performed with $^{137}$Cs, $^{88}$Y and Am/Be calibration sources, which were also used to verify the stability of the detector response during the 32 days  of the experiment. In addition, the energy resolution of the two C$_6$D$_6$ detectors was assessed from these calibration data.

The time-of-flight to energy calibration, that is the determination of the fixed flight path length, was based on the analysis of $^{197}$Au resonances, according to the method described in Ref. \cite{bibY}.

The well-established pulse height weighting technique, originally proposed by Maier-Leibnitz \cite{bibF}, was applied to the recorded signals. This procedure ensures  that the efficiency $\varepsilon_\gamma$ of a detector is independent of the de-excitation pattern of the capture $\gamma$-ray cascade. This condition is achieved by off-line modification of the detector response so that it becomes proportional to the detected $\gamma$-ray energy: $\varepsilon_\gamma=\alpha E_\gamma$. The weighting function $W(E)$ is determined by minimizing the expression:
\begin{linenomath}\begin{equation}
\sum_j\left[\int W(E')R(j;E')\mathrm{d}E'-\alpha E_\gamma(j)\right]^2
\end{equation}\end{linenomath}
with $R(j;E)$ as the detector response to a $\gamma$-ray of energy $E_\gamma(j)$. The spectra $R(j;E)$ were obtained by a detailed GEANT4 simulation. Within the simulation code the experimental setup of the two C$_6$D$_6$ detectors and the surrounding apparatus were described in detail. Thus the simulation takes into account the intrinsic and geometrical efficiency for detecting $\gamma$-rays as well as secondary effects such as absorption of $\gamma$-rays in the sample and the detection of photons scattered by the nearby components. Simulated monochromatic spectra were folded with the experimentally determined energy resolution of the two C$_6$D$_6$ detectors. The cut-off threshold of 200 keV applied during offline analysis of experimental data was also taken into account. The resulting spectra were used to calculate the weighting function, assumed to be a polynomial of 4th degree. A full description of the pulse height weighting technique adopted at n\_TOF may be found in Ref. \cite{bibG}.

\begin{figure}[t!]
\includegraphics[angle=90,width=1.0\linewidth,keepaspectratio]{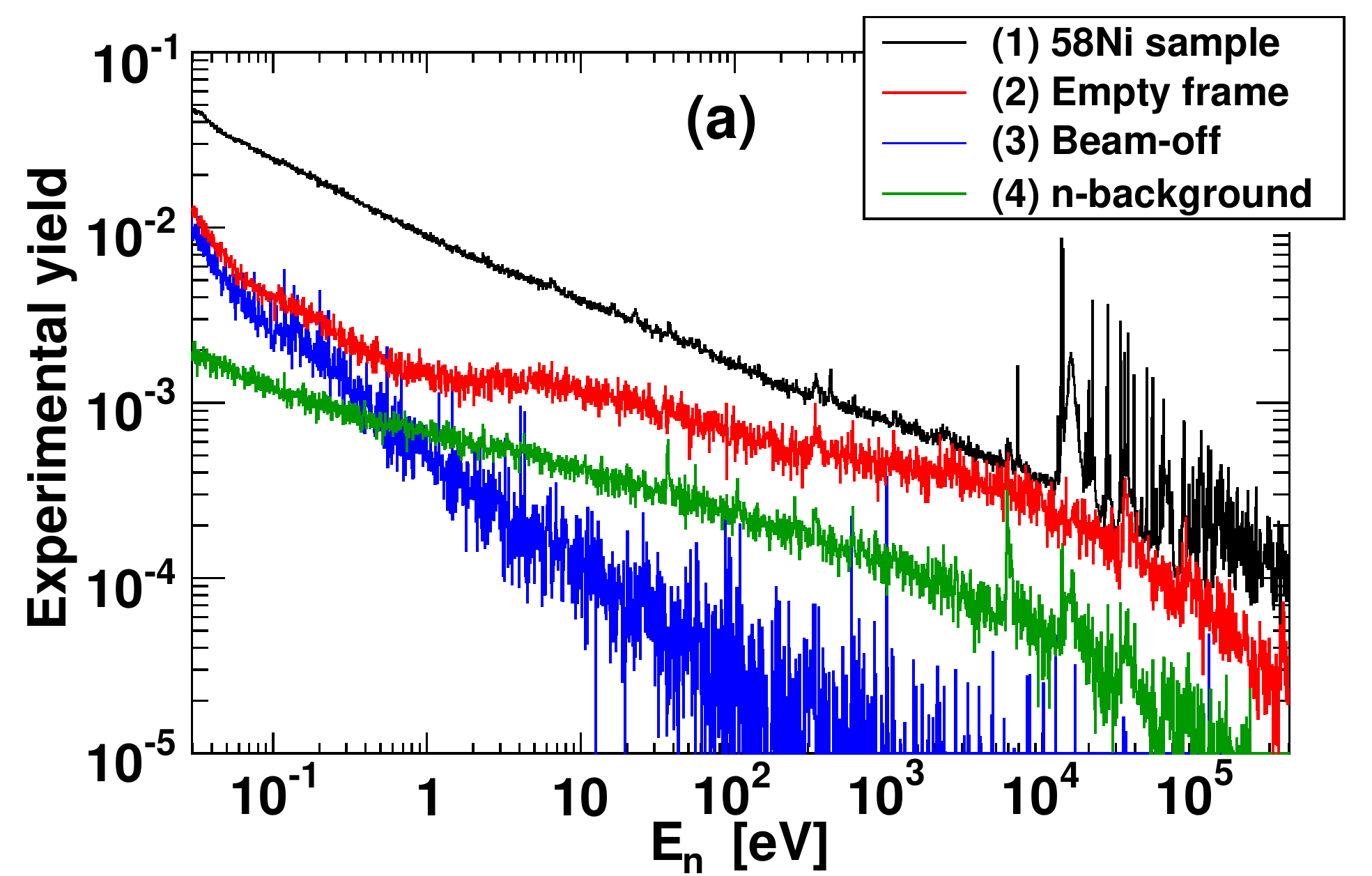}
\includegraphics[angle=90,width=1.0\linewidth,keepaspectratio]{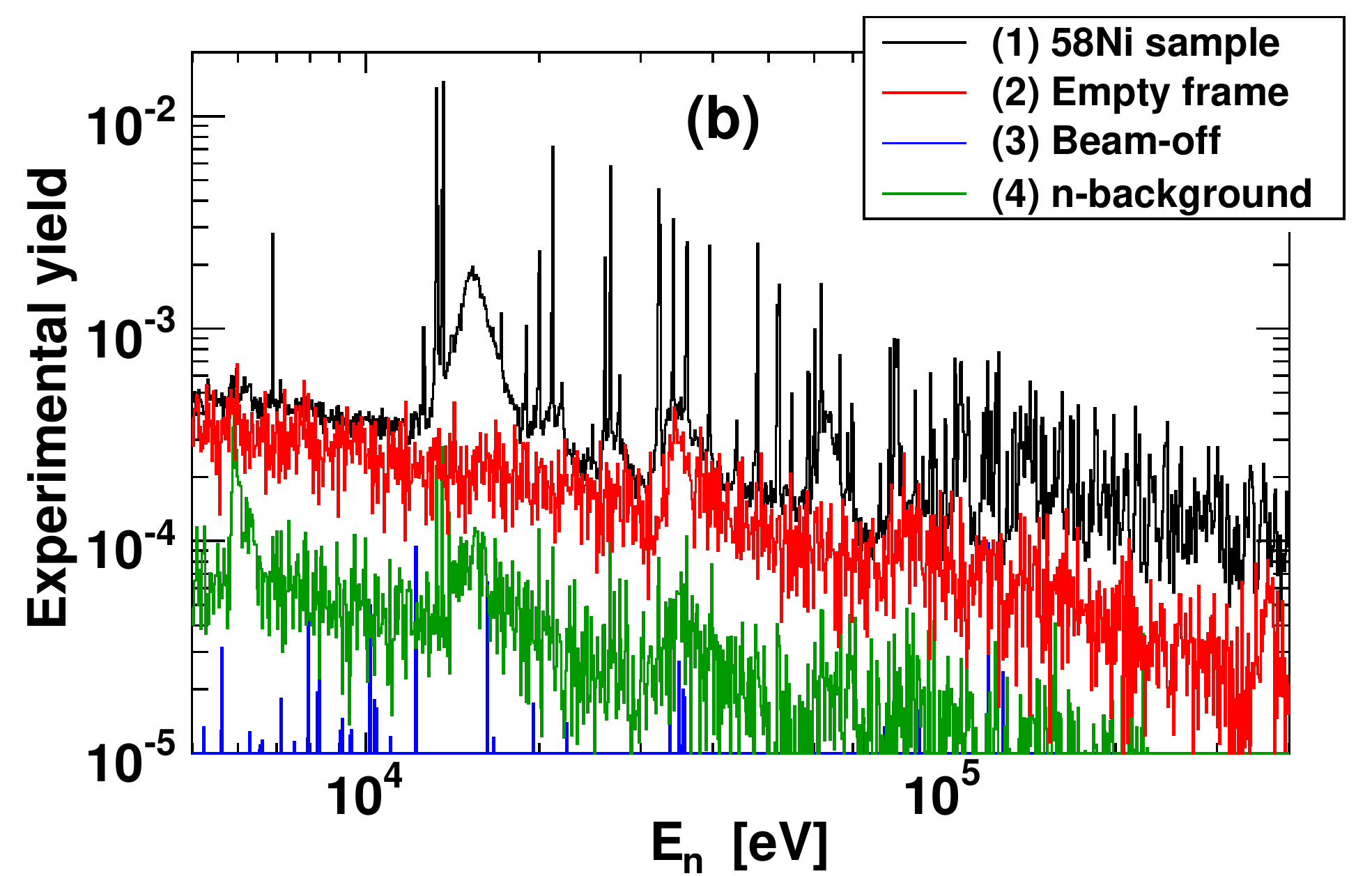}
\caption{(Color online) Top panel (a): total experimental yield of $^{58}$Ni (1) together with the background from an empty sample
holder (2), the ambient background (3), and the background from neutrons scattered off the sample (the neutron background) (4). Bottom panel (b): experimental yields within resonance-dominated region between 5 keV and 400 keV.}
\label{fig1}
\end{figure}

\begin{figure}[t!]
\includegraphics[angle=90,width=1.0\linewidth,keepaspectratio]{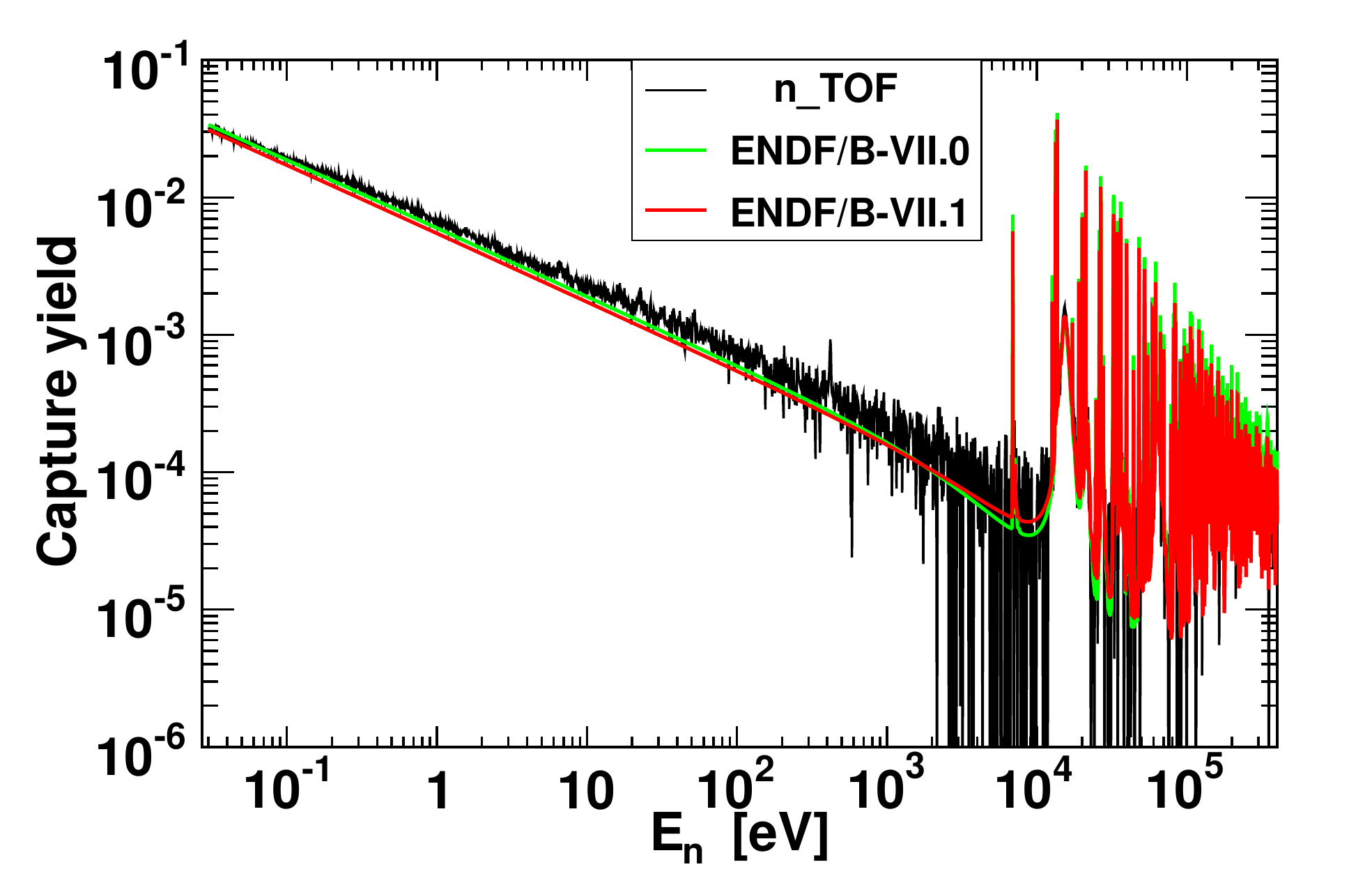}
\caption{(Color online) Capture yield of $^{58}$Ni -- corrected for an empty frame, ambient and the neutron backgrounds -- compared with the yield calculated from ENDF/B-VII.0 and ENDF/B-VII.1 resonance parameters. The n\_TOF resonances are mostly consistent with ENDF/B-VII.1 data.}
\label{fig2}
\end{figure}

Three main types of background had to be identified and subtracted. Figure \ref{fig1} shows the respective contributions to the total yield measured with the $^{58}$Ni sample. The first component, related to the neutron beam and independent of the sample, was determined with an empty sample frame, consisting of a 1.5 $\mu$m thick Mylar backing glued to the carbon fiber holder. Another component, caused by scattered in-beam $\gamma$-rays, was measured with a Pb sample, but was found to be negligible in this case. The ambient background -- caused by activation and natural radioactivity -- was measured  with the neutron beam turned off. Finally, the background caused by neutrons scattered off the sample itself -- referred to as the neutron background in Figs. \ref{fig1} and \ref{fig2}  -- was obtained for the first time at n\_TOF by dedicated GEANT4 simulation \cite{bibLL}. The approximate magnitude of the neutron background may in principle be estimated experimentally by inserting in the beam a sample of material, such as natural carbon, characterized by a very low capture-to-scattering ratio. Such a measurement, however, does not provide information on the time structure of neutron background. This problem can be circumvented by dedicated simulations, which give access to the wealth of information otherwise inaccessible by experiment, thus allowing for a more precise determination and subsequent subtraction of the neutron background. For the simulation GEANT4-9.6.p01 version was employed, which -- for the neutron-induced reactions -- relies largely on the tabulated cross section data from ENDF/B-VII.0. A detailed software replica of the experimental area and the materials inside it was implemented in the simulations. The energy deposited in C$_6$D$_6$ detectors, together with the corresponding time information was analyzed in the same manner as the experimental data. The reliability of the simulated neutron background was verified by comparing the GEANT4 results against the yield measured with a high-purity (99,95\%) carbon sample of 1 cm thickness and 2 cm diameter. A decade-wise comparison between the experimental and simulated yield is presented in Table \ref{tab5}. A good agreement is observed between the measured and the simulated yield for C in the whole neutron energy range of interest, from near-thermal to 400 keV. In particular, the results of the simulations reproduce very closely the measured C yield in the region of $^{58}$Ni resonances, i.e. above 1 keV, thus providing high confidence on the reliability of the GEANT4 simulations. On the other hand, a meaningful comparison cannot be performed below a few eV, since the measured C yield at low energy is heavily affected by the $\beta$-decay of $^{12}$B from the $^{12}$C($n,p$) reaction. Since  $^{12}$B has a half-life of 20 ms, the 6.35 MeV electrons from its decay can reach the C$_6$D$_6$ active volume within the same neutron bunch, $\sim$100 ms wide, at times corresponding to low reconstructed neutron energy (from thermal to a few eV). For this reason, the neutron background at low energy can only be reliably estimated from the simulations.

\begin{table}[b!]
\caption{Decade-wise comparison between the experimental and simulated yield for a $^\mathrm{nat}$C sample.}
\label{tab5}
\begin{tabular}{c>{\quad}c>{\quad}c}
\hline\hline
Energy&Experimental&GEANT4\\
range&yield&yield\\
\hline
$10^{-1}-10^{0}$ eV&$1.22\times10^{-2}$&$9.72\times10^{-3}$\\
$10^{0}-10^{1}$ eV&$5.23\times10^{-3}$&$4.43\times10^{-3}$\\
$10^{1}-10^{2}$ eV&$2.39\times10^{-3}$&$2.06\times10^{-3}$\\
$10^{2}-10^{3}$ eV&$1.14\times10^{-3}$&$1.00\times10^{-3}$\\
$10^{3}-10^{4}$ eV&$6.10\times10^{-4}$&$5.72\times10^{-4}$\\
$10^{4}-10^{5}$ eV&$2.39\times10^{-4}$&$2.21\times10^{-4}$\\
\hline\hline
\end{tabular}
\end{table}

All background components were properly normalized before subtracting them from the data measured with the $^{58}$Ni sample.

The capture yield as a function of neutron energy was calculated as:
\begin{linenomath}\begin{equation}
Y(E_n)=\frac{S_w(E_n)-B_w(E_n)}{N\cdot E_c(E_n)\cdot\phi(E_n)}
\end{equation}\end{linenomath}
where $S_w(E_n)$ and $B_w(E_n)$ are the total sample related and background counts, respectively, to which a weighting function has been applied, making the efficiency to detect a capture event proportional to the excitation energy $E_c$ of a compound nucleus ($E_c$ being related to the neutron separation energy $S_n=8.99$ ~MeV of $^{59}$Ni). Since the sample dimension was smaller than the spatial beam profile, $\phi(E_n)$ is the neutron flux intercepted by the sample. It was obtained by multiplying the evaluated neutron flux $\Phi(E_n)$, described in Ref. \cite{bibL}, by the simulated energy dependent beam interception factor \cite{bibB}. The absolute yield normalization factor $N$, accounting for several experimental effects, was evaluated by means of the saturated resonance technique, applied to the 4.9 eV resonance of $^{197}$Au \cite{bibH}.

Figure \ref{fig2} shows the yield measured at n\_TOF, compared with the one calculated on the basis of ENDF/B-VII.0 and ENDF/B-VII.1 resonance parameters. The evaluation from ENDF/B-VII.0 is also representative of other major libraries such as JENDL-4.0, JEFF-3.1.2, CENDL-3.1 and ROSFOND-2010 \cite{bibUU}. The difference between two consecutive versions of the ENDF library and their comparison with the n\_TOF data will be discussed below.

\section{Data Analysis}

\subsection{Thermal point}

\begin{figure}[b!] 
\includegraphics[angle=90,width=1.0\linewidth,keepaspectratio]{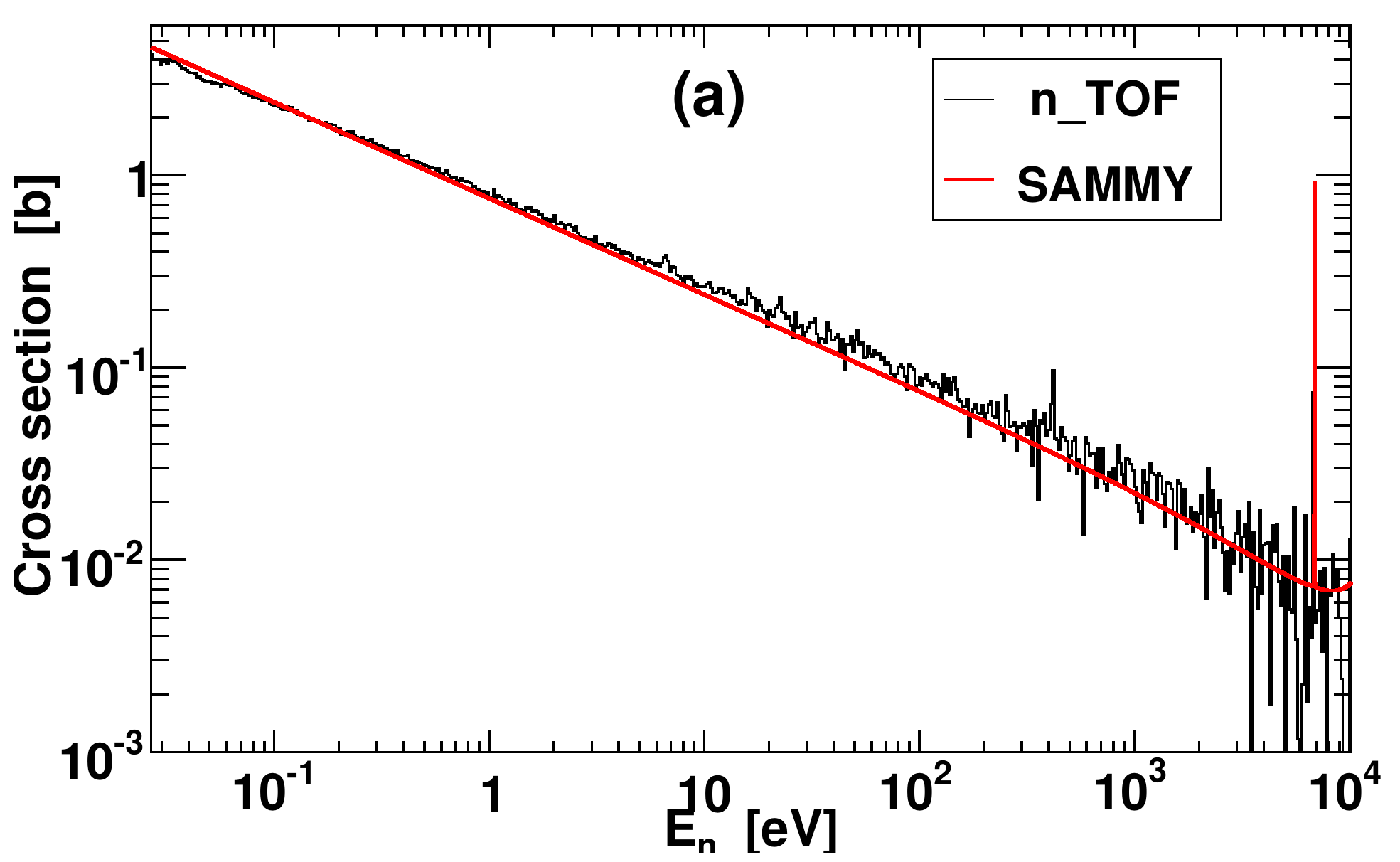}
\includegraphics[angle=90,width=1.0\linewidth,keepaspectratio]{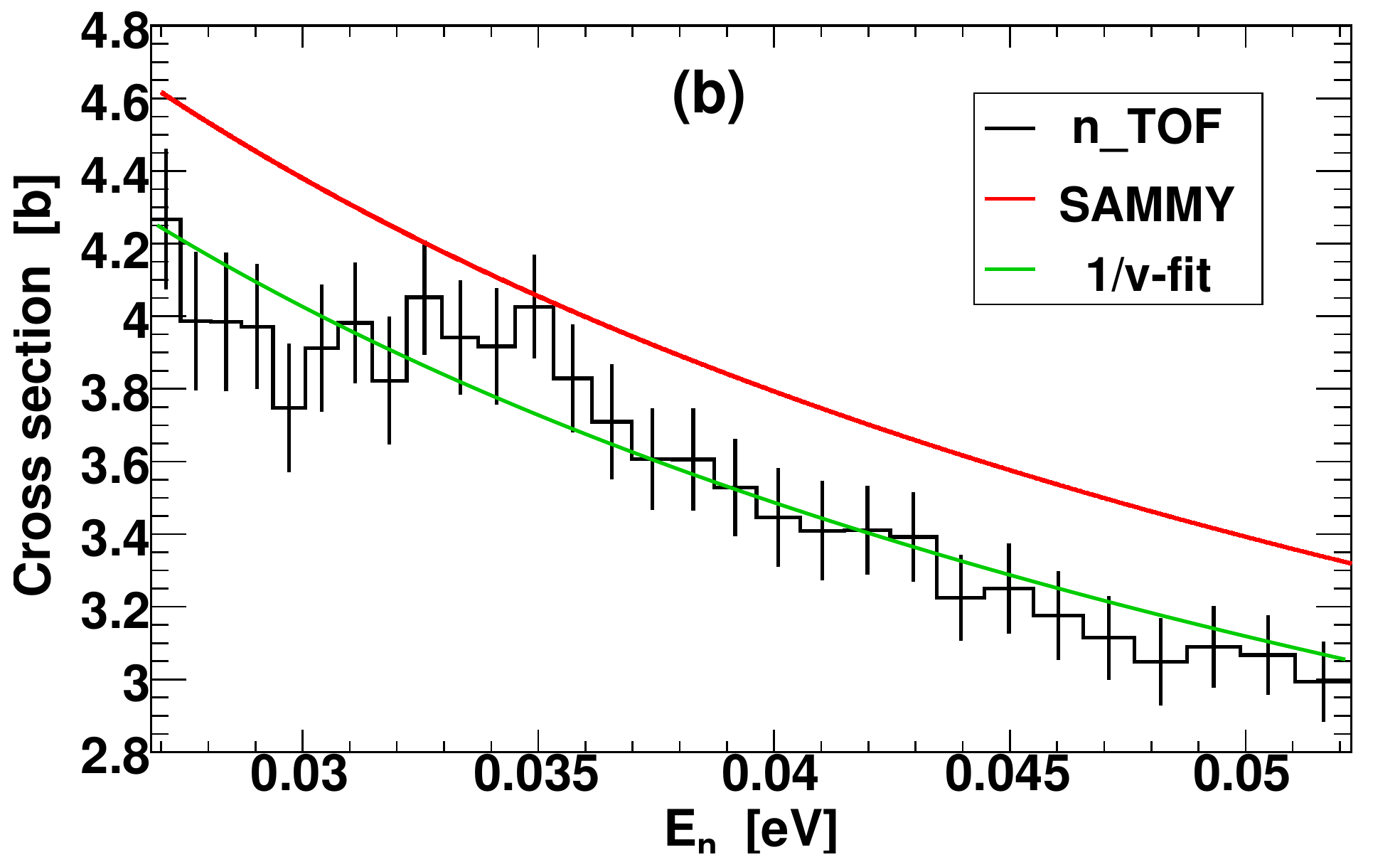}
\caption{(Color online) Top panel (a): cross section from 27~meV to 10 keV neutron energy, compared to the SAMMY fit with two negative-energy resonances. Bottom panel (b): comparison between the global SAMMY fit and an independent 1/$v$-fit to the data between 27 meV and 50 meV.}
\label{fig8}
\end{figure}

\begin{table}[t!]
\caption{Comparison of thermal $^{58}$Ni cross section with previous experimental data from EXFOR \cite{bibO}, the recommended value of Ref. \cite{bibQQ} and data from the ENDF/B library \cite{bibSS,bibTT}.}
\label{tab3}
\begin{tabular}{l<{\quad}>{\quad}c}
\hline\hline
\textbf{Source}&\textbf{$\boldsymbol{\sigma}$(25.3 meV)}\\
\hline
n\_TOF&$4385\pm22_\mathrm{stat}\pm137_\mathrm{sys}$ mb\\\\
Pomerance\footnotemark[1] (1952)&$4200\pm336$ mb\\
Ishaq\footnotemark[1] (1977)&4500 mb\\
Carbonari (1988)&$4520\pm100$ mb\\
Weselka\footnotemark[1] (1991)&$4600\pm300$ mb\\
Venturini\footnotemark[1] (1997)&$4400\pm200$ mb\\
Raman\footnotemark[1] (2004)&$4130\pm50$ mb\\\\
ENDF/B-VII.0&4621 mb\\
ENDF/B-VII.1&4227 mb\\
Mughabghab&4370 mb\\
\hline\hline
\end{tabular}
\footnotetext[1]{Data reported as Maxwellian average.}
\end{table}

The experimental data were fitted by a multi-level R-matrix code SAMMY \cite{bibI}, which also accounts for the resolution function of the neutron beam, Doppler broadening of capture resonances, multiple scattering and self-shielding effects. A reasonable fit for energies up to 10 keV was obtained by considering two resonances with negative energy, as adopted in the latest versions of the evaluated data libraries, as well as suggested by Mughabghab in his latest compilation \cite{bibQQ}. In combination with the transmission data, the negative resonances may be used for characterization of the bound states of a nucleus. The energy and neutron width of the two resonances were fixed at the value adopted from the ENDF/B-VII.1 parameters database -- accessed via JANIS interface \cite{bibPP} -- while the capture width was left free. The following parameters were found to best fit the n\_TOF data:
\begin{linenomath}\begin{align*}
\begin{split}
(E_R^{(1)},\Gamma_\gamma^{(1)},\Gamma_n^{(1)})&=(-78318\:\mathrm{eV},25\:\mathrm{eV},40685\:\mathrm{eV})\\
(E_R^{(2)},\Gamma_\gamma^{(2)},\Gamma_n^{(2)})&=(-11674\:\mathrm{eV},1.6\:\mathrm{eV},4262.7\:\mathrm{eV})
\end{split}
\end{align*}\end{linenomath}
The result of the fit is shown in Figure \ref{fig8}. Since the fit slightly overestimates the yield at low energy, it was not used to extract the cross section at thermal energy (25.3~meV). To this end, a different method was used instead, consisting in fitting the cross section with an $1/v$ behavior only in the region between 27 meV and 50 meV, and extrapolating the fit to the neutron energy of 25.3 meV. The result of such a fit is shown in the panel (b) of Figure \ref{fig8} by the green line. The value extracted in this way for the thermal cross section at 25.3 meV is $4385\pm22_\mathrm{stat}\pm137_\mathrm{sys}$ mb, close to the recent result from Raman \emph{et al.} \cite{bibMM}, and in agreement within 0.4\% with the recommended value of 4370 mb from Mughabghab \cite{bibQQ}. The systematic uncertainty originates from the uncertainties on the neutron flux (2\% at thermal energies), weighting function (2\%) and the beam interception factor (1.3\%). The statistical error was determined from the uncertainty on the fit parameters. The significance of the thermal value is related to nickel being an important structural material of interest for current and future nuclear reactors. A comparison with previous measured and evaluated thermal values is listed in Table \ref{tab3}, where the n\_TOF results are compared against evaluations from ENDF/B-VII.0 and ENDF/B-VII.1. In addition, the available experimental data from the EXFOR database \cite{bibO} are listed, most of them reported as Maxwellian averages at 25.3~meV. We remind the reader that, for a pure $1/v$ dependence of the capture cross section, the Maxwellian average corresponds to $\sigma(kT)$. At thermal energies the major modifications in ENDF/B-VII.1 are mostly related to the results from Raman \emph{et al.} \cite{bibMM}, who reported a significantly lower capture cross section, relative to the previous version of ENDF. In conclusion, our results at thermal energies confirm the latest ENDF/B-VII.1 evaluation, where the cross section has been reduced by approximately 10\% with respect to all other previous evaluations (ENDF/B-VII.0, JENDL-4.0, JEFF-3.1.2, CENDL-3.1, ROSFOND-2010).

\begin{figure}[b!]
\includegraphics[width=1.0\linewidth,keepaspectratio]{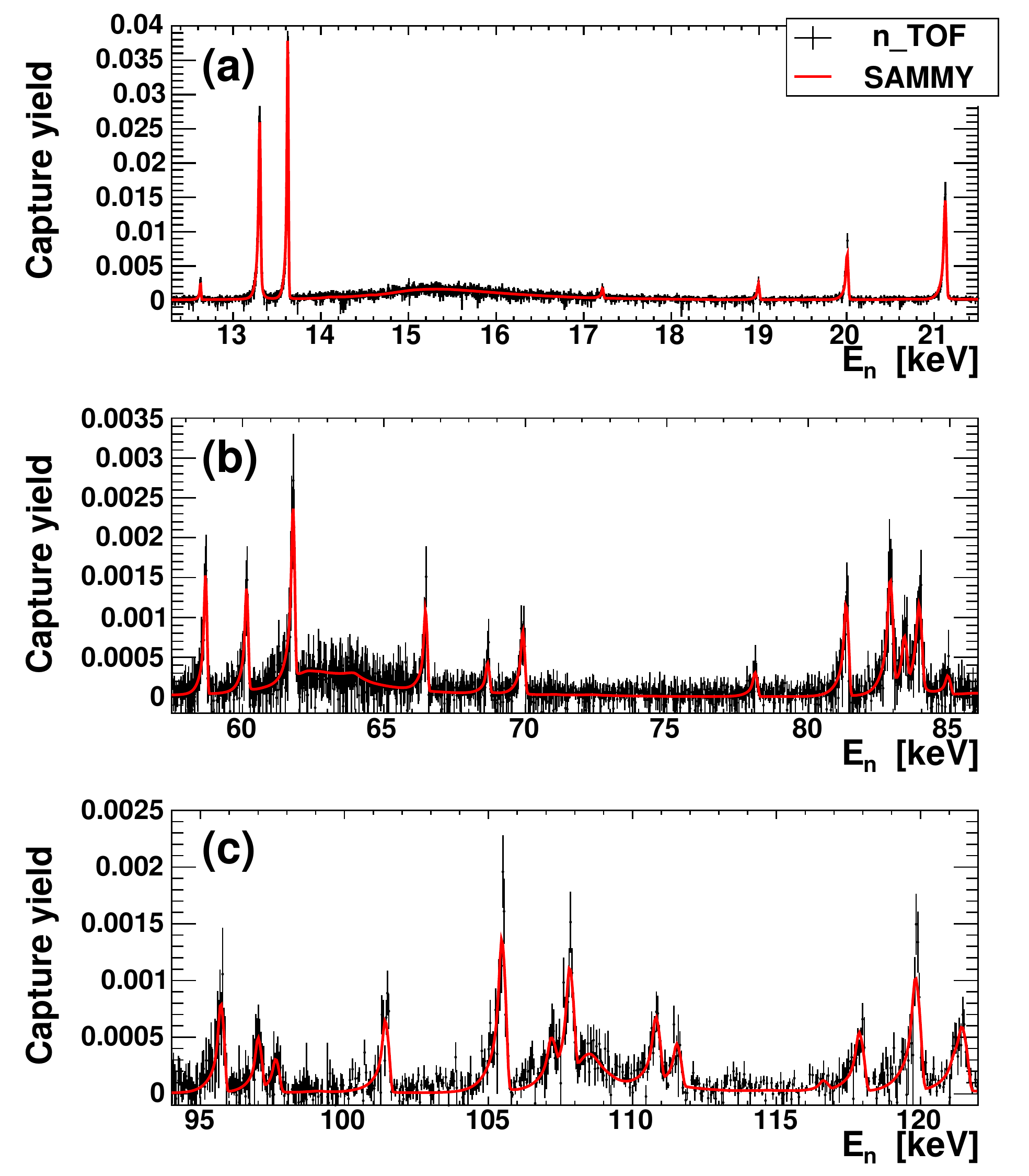}
\caption{(Color online) Examples of SAMMY fits of several resonances in the capture yield of $^{58}$Ni.}
\label{fig3}
\end{figure}

\begin{table}[b!]
\caption{List of 51 resolved neutron capture resonances of $^{58}$Ni up to 122 keV. The resonance energy $E_R$ and capture kernel $K$ are listed together with orbital angular momentum $\ell$, resonance spin $J$ and the statistical spin factor $g_s$. The values of $\ell$ and $J$ have been adopted from ENDF/B-VII.1 parameters database.}
\label{tab2}
\begin{tabular}{c>{\qquad}c>{\qquad}c>{\qquad}c>{\qquad}c}
\hline
\hline
$\boldsymbol{E_R}$ \textbf{(keV)}&$\boldsymbol{\ell}$&$\boldsymbol{J}$&$\boldsymbol{g_s}$&$\boldsymbol{K}$ \textbf{(meV)}\\
\hline
6.8927(6)&1&1/2&1&$20\pm2_\mathrm{stat}\pm1_\mathrm{sys}$\\
12.616(1)&1&1/2&1&$25\pm4\pm1$\\
13.2927(6)&1&1/2&1&$601\pm21\pm33$\\
13.6114(3)&1&3/2&2&$591\pm17\pm33$\\
15.350(7)&0&1/2&1&$1279\pm41\pm71$\\
17.200(2)&1&1/2&1&$38\pm9\pm2$\\
18.976(2)&2&5/2&3&$71\pm10\pm4$\\
19.987(1)&1&1/2&1&$256\pm18\pm14$\\
21.1051(8)&1&3/2&2&$663\pm27\pm37$\\
26.024(2)&1&3/2&2&$263\pm25\pm15$\\
26.596(1)&1&3/2&2&$864\pm40\pm48$\\
27.573(4)&1&1/2&1&$33\pm14\pm2$\\
32.207(3)&2&5/2&3&$388\pm39\pm21$\\
32.337(3)&1&1/2&1&$1120\pm74\pm62$\\
34.178(3)&1&3/2&2&$689\pm67\pm38$\\
36.073(3)&0&1/2&1&$1020\pm90\pm56$\\
39.492(3)&2&3/2&2&$719\pm62\pm40$\\
43.952(4)&2&5/2&3&$123\pm26\pm7$\\
47.822(3)&1&3/2&2&$1128\pm87\pm62$\\
51.839(5)&2&5/2&3&$780\pm87\pm43$\\
52.142(4)&2&3/2&2&$830\pm80\pm46$\\
54.711(3)&2&3/2&2&$235\pm44\pm13$\\
58.617(4)&1&3/2&2&$570\pm62\pm32$\\
60.067(7)&1&3/2&2&$606\pm79\pm34$\\
61.706(7)&1&1/2&1&$1115\pm123\pm62$\\
62.7(2)&0&1/2&1&$2621\pm248\pm145$\\
66.379(6)&1&3/2&2&$565\pm78\pm31$\\
68.5706(8)&2&3/2&2&$205\pm53\pm11$\\
69.813(9)&1&1/2&1&$529\pm76\pm29$\\
78.006(6)&1&1/2&1&$238\pm57\pm13$\\
81.217(1)&2&3/2&2&$1021\pm146\pm48$\\
82.77(1)&1&3/2&2&$1532\pm193\pm71$\\
83.268(8)&0&1/2&1&$571\pm133\pm27$\\
83.79(1)&1&1/2&1&$1278\pm169\pm59$\\
84.803(4)&2&3/2&2&$234\pm77\pm11$\\
89.832(9)&1&3/2&2&$705\pm128\pm33$\\
95.5580(5)&2&5/2&3&$1025\pm175\pm48$\\
96.850(6)&2&5/2&3&$606\pm137\pm28$\\
97.46(2)&1&1/2&1&$434\pm93\pm20$\\
101.255(7)&2&5/2&3&$1011\pm98\pm47$\\
105.294(7)&2&3/2&2&$2320\pm144\pm108$\\
106.993(7)&1&1/2&1&$506\pm105\pm23$\\
107.640(8)&2&3/2&2&$1549\pm156\pm72$\\
108.45(8)&0&1/2&1&$2256\pm232\pm105$\\
110.627(7)&1&3/2&2&$993\pm113\pm46$\\
111.3547(2)&2&5/2&3&$697\pm31\pm32$\\
116.420(4)&1&1/2&1&$180\pm58\pm8$\\
117.67(1)&1&3/2&2&$1185\pm116\pm55$\\
119.624(8)&2&5/2&3&$2338\pm211\pm108$\\
120.958(8)&1&1/2&1&$332\pm118\pm15$\\
121.25(1)&2&3/2&2&$1377\pm163\pm64$\\
\hline
\hline
\end{tabular}
\end{table}

\begin{figure}[b!]
\includegraphics[angle=90,width=1.0\linewidth,keepaspectratio]{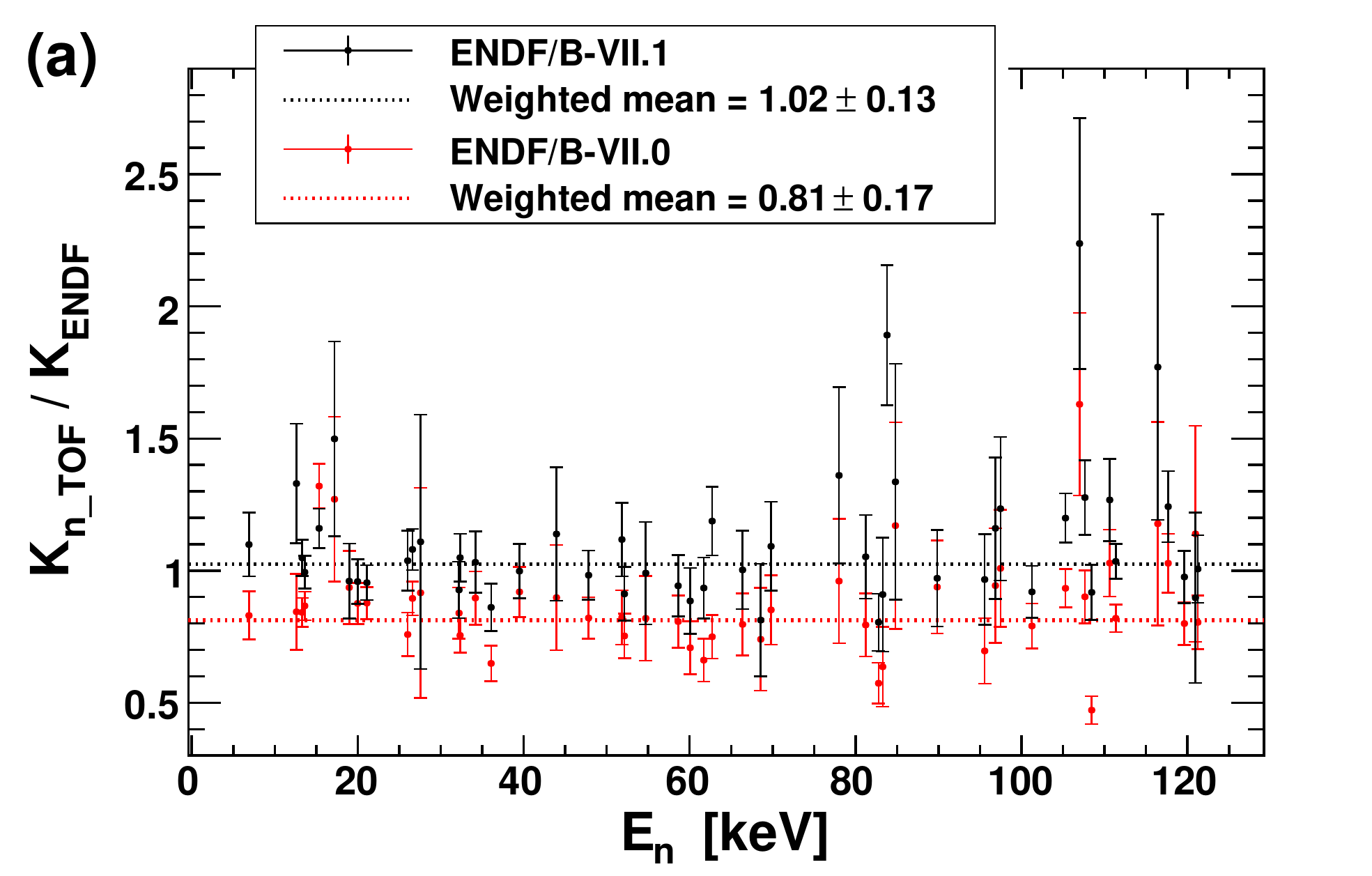}
\includegraphics[angle=90,width=1.0\linewidth,keepaspectratio]{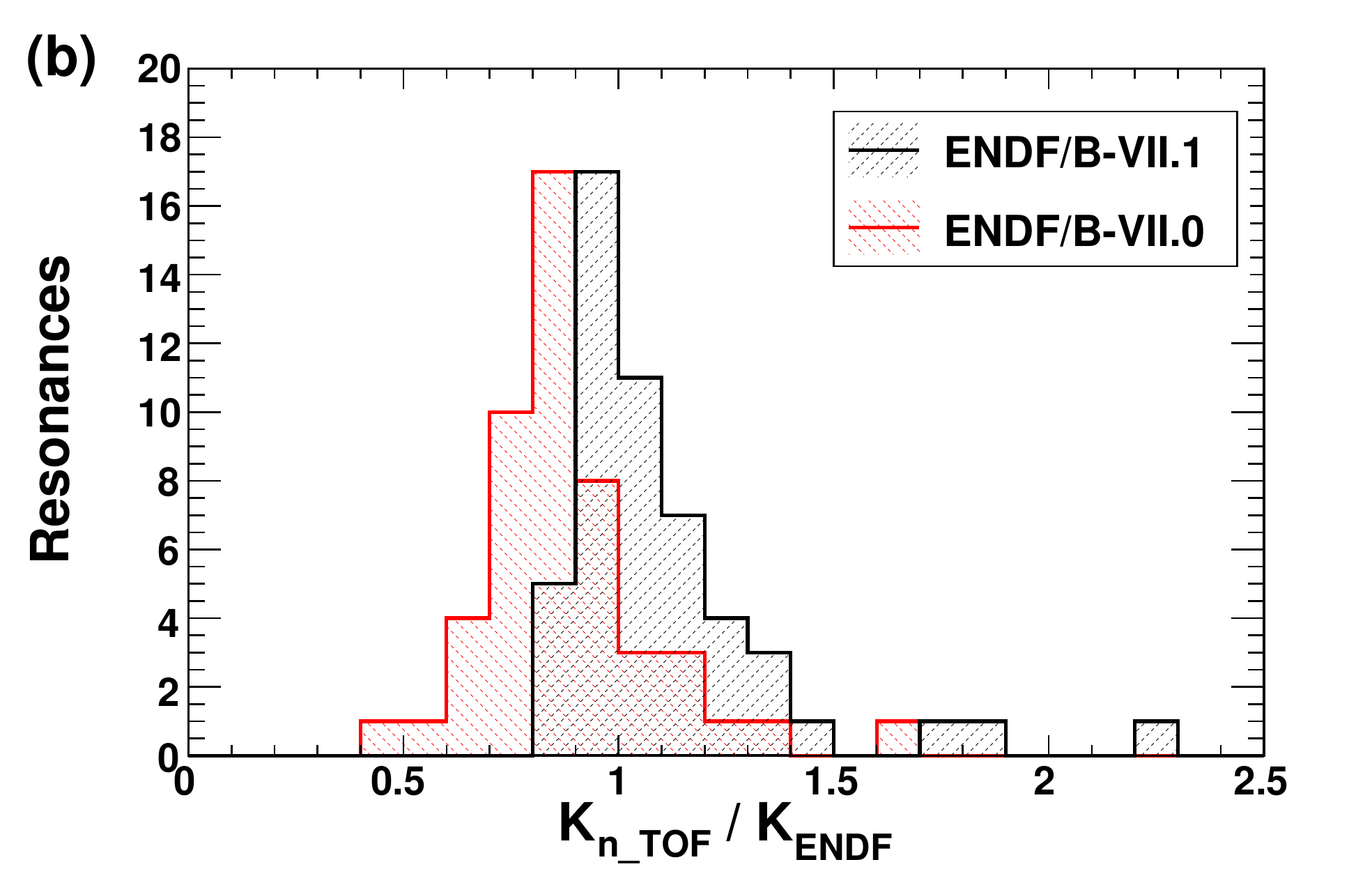}
\caption{(Color online) Top panel (a): kernel ratios: n\_TOF vs. ENDF/B-VII.0 and ENDF/B-VII.1. The weighted means of ratios are also indicated. Bottom panel (b): distribution of kernel ratios from top panel.}
\label{fig4}
\end{figure}

\subsection{Resolved resonance region}

Up to 122 keV neutron energy, a total of 51 resolved resonances were identified and analyzed. Figure \ref{fig3} presents an example of several resonances fitted by SAMMY. The associated capture kernels:
\begin{linenomath}\begin{equation}
K=g_s\frac{\Gamma_n\Gamma_\gamma}{\Gamma_n+\Gamma_\gamma}
\end{equation}\end{linenomath}
were calculated from the neutron $\Gamma_n$ and radiative capture $\Gamma_\gamma$ widths. For the neutron spin $I_n=1/2$ and the $^{58}$Ni ground state spin $I_\mathrm{^{58}Ni}=0$, the statistical spin factor $g_s$:
\begin{linenomath}\begin{equation}
g_s=\frac{2J+1}{(2I_n+1)(2I_\mathrm{^{58}Ni}+1)}
\end{equation}\end{linenomath}
depends only on the resonance spin $J$. The orbital angular momentum $\ell$, resonance spin $J$ and initial values for the resonance widths $\Gamma_n$ and $\Gamma_\gamma$  were adopted from the ENDF/B-VII.1 parameters database \cite{bibTT}. Out of 56 resonances listed therein up to 122 keV, 5 could not be resolved, i.e. distinguished from the fluctuations in the baseline of the n\_TOF yield due to the very small associated kernels. These are the resonances at 24.77 keV, 35.06 keV, 48.47 keV, 83.97 keV and 92.73 keV. In addition, no new resonances were found. In the course of the fitting procedure the largest of the widths was kept fixed -- with exception of cases when $\Gamma_n$ and $\Gamma_\gamma$ were comparable -- while the smaller one was left free in order to accurately reproduce the resonance shapes. The sources of the systematic uncertainty are the neutron flux (2\% for the flux within 27 meV -- 200 eV, 3\% within 200 eV -- 8 keV, 5\% within 8 keV -- 80 keV, 4\% within 80 keV -- 1 MeV), weighting function (2\%) and the beam interception factor (1.3\%).

Table \ref{tab2} lists the capture kernels for all 51 resolved resonances. Together with those, the already mentioned  resonances at negative energy were included in the global SAMMY fit, to account for the $1/v$ dependence of the cross section at low energy, which is partially related -- as discussed later -- with s-wave Direct or Direct-Semidirect Capture \cite{bibNN} components. Figure \ref{fig4} shows the ratios between kernels determined in this work and those evaluated from ENDF/B-VII.0 and ENDF/B-VII.1. A weighted mean of $0.81\pm0.17$ is obtained  in case of ENDF/B-VII.0, while for ENDF/B-VII.1 the weighted mean of $1.02\pm0.13$ indicates an overall agreement within 2\% with the latest evaluation. The distributions of kernel ratios are also shown.

\subsection{Unresolved resonance region}

\begin{figure}[b!]
\vspace*{-2mm}
\includegraphics[angle=90,width=1.0\linewidth,keepaspectratio]{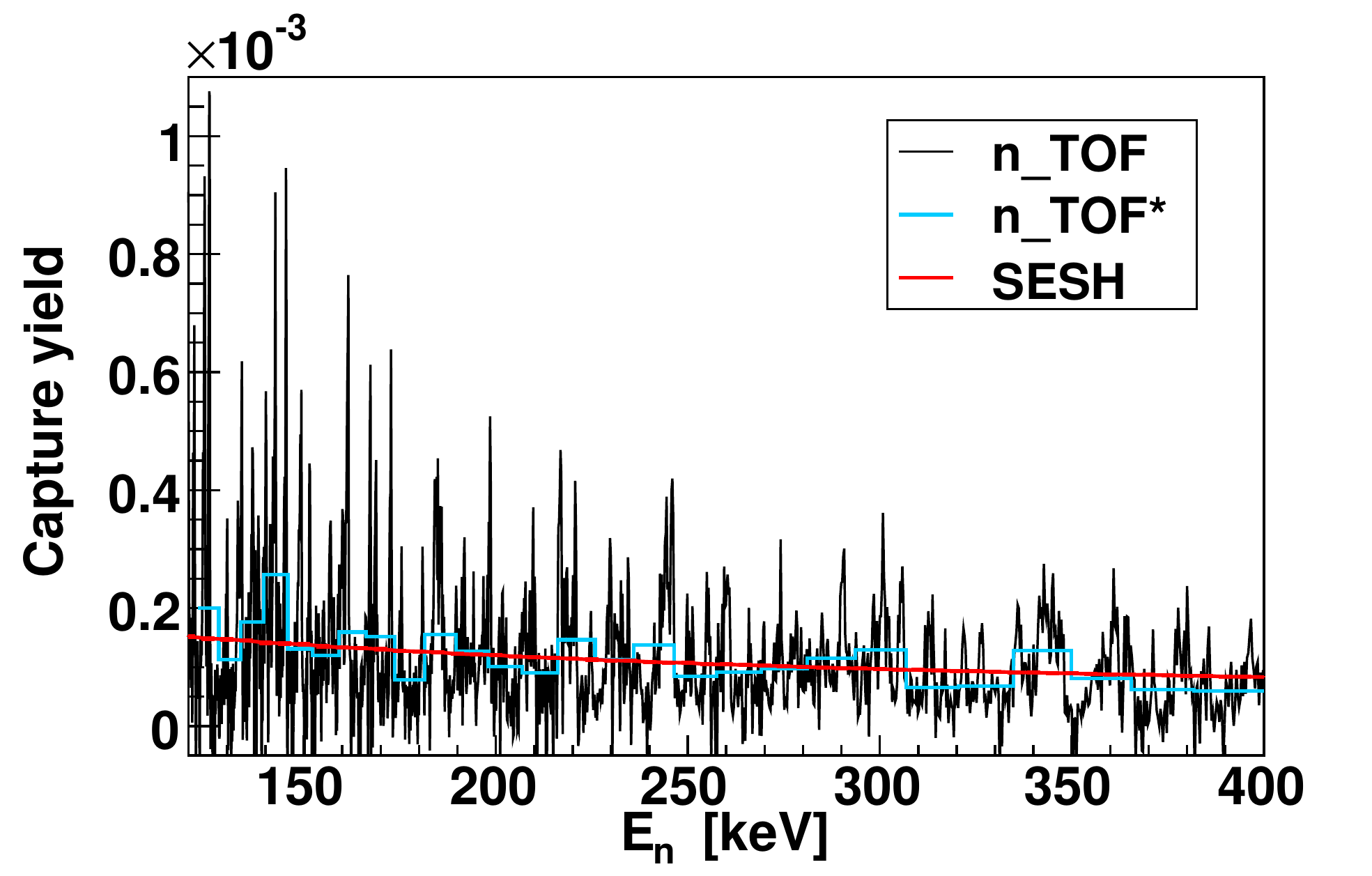}
\caption{(Color online) Measured yield compared to the yield simulated by SESH code. While the two plots designated as n\_TOF and n\_TOF$^*$ show the same data, the n\_TOF$^*$ plot is shown in 50 times coarser binning in order to reproduce the average yield behavior and to facilitate the visual comparison with SESH results.}
\label{fig9}
\end{figure}

The energy range above 122 keV was treated as an unresolved resonance region, using the code SESH \cite{bibJ} to extract the capture cross section $\sigma(E_n)$ up to $E_n~=~1$~MeV. Although a resonant structure is still evident in the yield up to 400 keV, as shown in Fig. \ref{fig9} (black histogram), this region was treated as unresolved due to the degradation in the energy resolution. The code SESH was used in order to reconstruct the smoothed capture cross section within this range. The set of optimized SESH parameters could then be used to extend the neutron capture cross section up to 1 MeV. The yield obtained with optimized parameters is also shown in Fig. \ref{fig9} (red curve). The good reproduction of the average behaviour of the cross section obtained with the SESH fit is demonstrated by the comparison with the capture yield averaged over a wider energy bin, also shown in the figure (blue histogram).

\section{Maxwellian averaged cross sections}

\begin{table*}[t!]
\caption{List of MACS determined in this work, compared to the values from KADoNIS v0.3 and to the two consecutive versions of ENDF.}
\label{tab4}
\begin{tabular}{cc>{\quad}c>{\qquad}c>{\qquad}c}
\hline\hline
\multirow{2}{*}{$\boldsymbol{kT}$ \textbf{(keV)}}&\multicolumn{4}{c}{\textbf{MACS (mb)}}\\
\cline{2-5}
&\textbf{n\_TOF}&\textbf{KADoNIS v0.3}&\textbf{ENDF/B-VII.0}&\textbf{ENDF/B-VII.1}\\
\hline
5&$41.3\pm0.6_\mathrm{stat}\pm2.3_\mathrm{sys}$&38.3&39.8&38.2\\
10&$50.1\pm0.7\pm2.8$&50.1&52.0&48.1\\
15&$45.9\pm0.7\pm2.5$&48.1&49.9&44.9\\
20&$41.0\pm0.6\pm2.2$&44.5&46.2&40.6\\
25&$37.2\pm0.6\pm2.0$&41.3&42.9&37.0\\
30&$34.2\pm0.6\pm1.8$&38.7&40.2&34.1\\
40&$30.3\pm0.5\pm1.5$&35.0&36.3&30.0\\
50&$27.7\pm0.4\pm1.4$&32.3&33.5&27.1\\
60&$25.8\pm0.3\pm1.3$&30.1&31.4&25.0\\
80&$23.2\pm0.3\pm1.1$&27.0&28.0&21.9\\
100&$21.3\pm0.2\pm1.0$&24.4&25.4&19.7\\
\hline
\hline
\end{tabular}
\end{table*}

\begin{figure}[b!]
\includegraphics[width=1.0\linewidth,keepaspectratio]{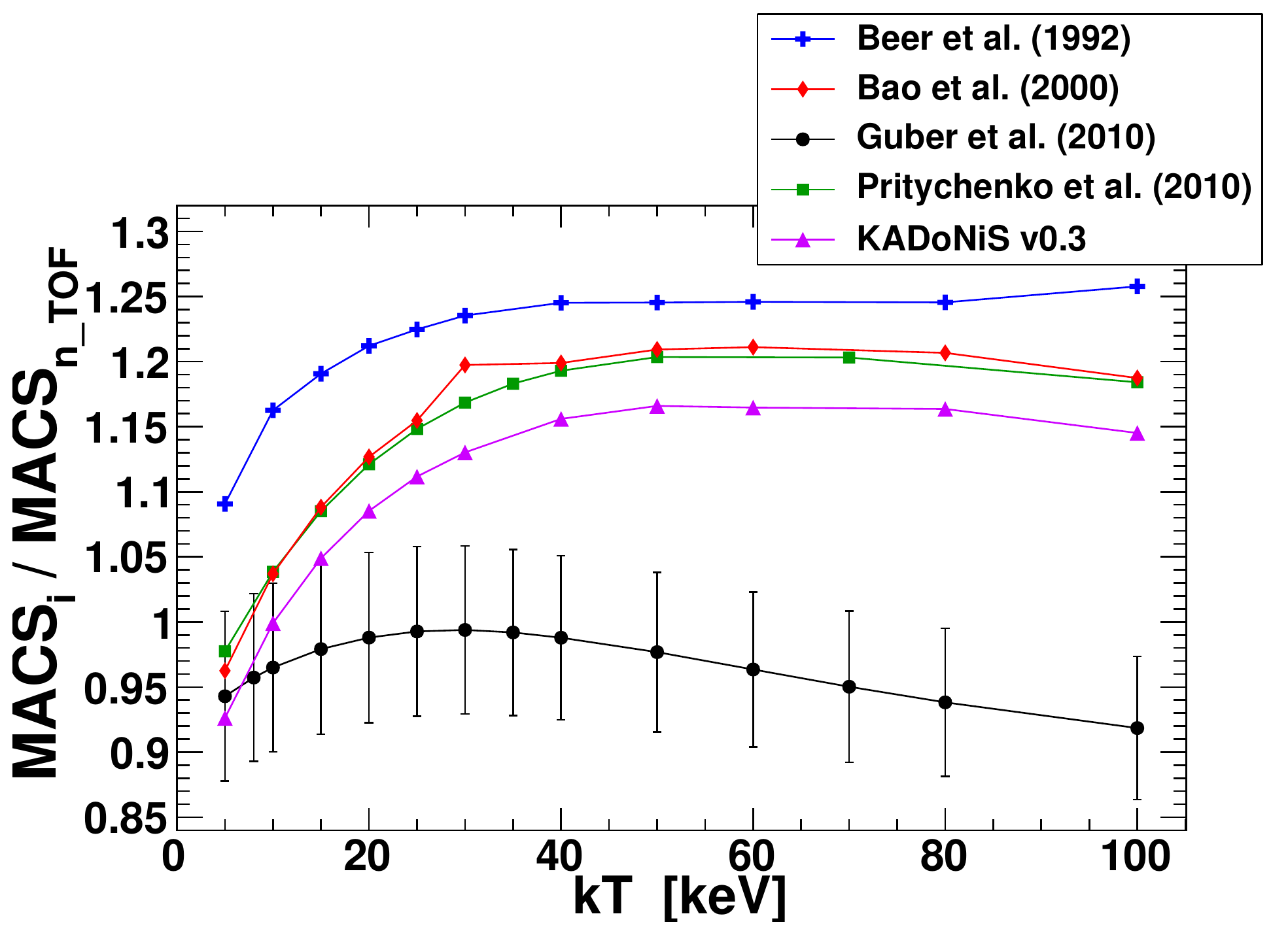}
\caption{(Color online) MACS values from several sources \cite{ bibK,bibN,bibP,bibR,bibS} relative to the n\_TOF data.}
\label{fig6}
\end{figure}

Using the resonance parameters obtained by SAMMY fits below 122 keV and the cross section from SESH above 122 keV, Maxwellian averaged cross sections (MACS):
\begin{linenomath}\begin{equation}
\langle\sigma\rangle_{kT}=\frac{2}{\sqrt{\pi}}\cdot\frac{1}{(kT)^2}\int_0^\infty\sigma(E_n)E_ne^{-E_n/kT}\mathrm{d}E_n
\end{equation}\end{linenomath}
have been determined over the full range of temperatures relevant for astrophysical purposes. An important remark concerns the presence of a Direct Capture (DC) and Direct-Semidirect (DSD) Capture \cite{bibNN} component. Theoretical calculations \cite{bibN} predict that such a component plays an important role at low energy in this isotope. In particular, s-wave neutrons, characterized by an energy dependence close to $1/v$ behavior, may account for more than 30\% of the thermal cross section value, while a much smaller contribution is expected for p-wave neutrons at higher energy. It should be considered that the low energy DC and DSD components can be accounted for by the negative energy resonances, as essentially done in the evaluations. In the present work, since the capture yield was determined with good accuracy down to thermal neutron energy, and a reasonable global fit was obtained from thermal energy to a few keV, it was not necessary to consider in the MACS an additional contribution of the s-wave DC or DSD capture component, calculated from theoretical models, as done in Ref. \cite{bibN}. We remark that, although conceptually different, the two methods used for the determination of this non-resonant component in the MACS are practically equivalent. The only missing correction regards the DC component of p-wave neutrons. The calculations based on the theoretical model from Ref. \cite{bibBBB} indicate that p-wave neutrons contribute less than 1\% to the extracted MACS at all temperatures.

\begin{figure}[b!]
\includegraphics[angle=90,width=1.0\linewidth,keepaspectratio]{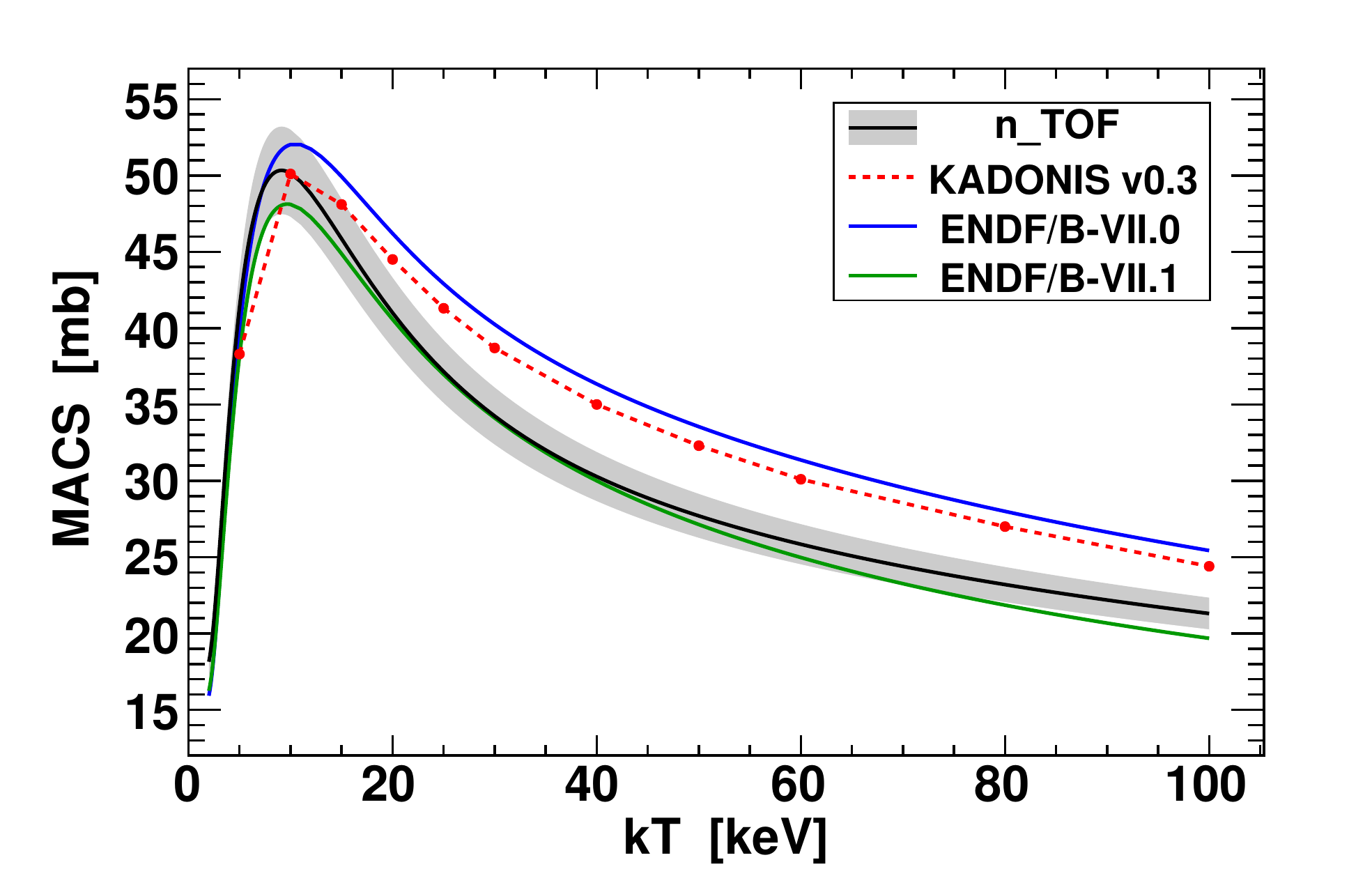}
\caption{(Color online) Compilation of MACS over the temperature range of astrophysical importance. The shaded region represents the uncertainty range for the n\_TOF data.}
\label{fig5}
\end{figure}

Figure \ref{fig6} shows the MACS from several earlier works \cite{bibK,bibN,bibP,bibR,bibS} relative to the n\_TOF data. The present results confirm Guber's results around $kT$ = 30 keV, a temperature domain of special importance for the stellar nucleosynthesis. However, for temperatures above 70 keV the n\_TOF results show a deviation larger than 5\% relative to Guber's data.

Figure \ref{fig5} shows the present MACS compared to those evaluated from ENDF/B-VII.0 and ENDF/B-VII.1 (the values of the MACS are also reported in Table \ref{tab4}). Below 10 keV the n\_TOF results are in better agreement with MACS values calculated from ENDF/B-VII.0, which may be confirmed from Table \ref{tab4}. In a 15 keV -- 50 keV range the latest ENDF/B-VII.1 evaluation is clearly favored. Above 50 keV the n\_TOF results lie between the values predicted by both libraries. Since all other major libraries essentially reproduce the evaluations in ENDF/B-VII.0, the n\_TOF results call for a revision of the $^{58}$Ni capture data in those data libraries. In Fig.~\ref{fig5} the available MACSs from KADoNIS v0.3 compilation \cite{bibK} are also shown. The plotted data are listed in Table \ref{tab4} for all $kT$ values found in KADoNIS v0.3 database. Both the statistical and systematic uncertainty in the MACS have been obtained by error propagation, starting from the values reported for the resolved and unresolved resonance region. Systematic uncertainties are the same as those for the systematic uncertainty of kernels.

The impact of our new results on the $s$-process was studied with a full 25M$_{\odot}$ stellar model with an initial metal content of $Z$ =0.02 \cite{bibOO}. The $s$-process nucleosynthesis is provided by the post-processing NuGrid code MPPNP \cite{bibW}. The final abundance distribution exhibits an effect that is essentially limited to the abundance of $^{58}$Ni itself. Due to the smaller MACS values compared to KADoNiS v0.3, $^{58}$Ni is less efficiently depleted, leading to a 60\% higher final abundance in the $s$-processed material. However, the overall effect of the new $^{58}$Ni values on the $s$-process abundance pattern is marginal, mostly because the $^{58}$Ni abundance is much smaller (4.3\% \cite{bibCCC}) compared to the dominant $^{56}$Fe seed.

\section{Conclusions}

The radiative neutron capture cross section of $^{58}$Ni has been measured in a wide energy range from 27 meV up to 400 keV, taking advantage of the high instantaneous neutron flux available at the n\_TOF facility at CERN, together with the low neutron sensitivity of two liquid C$_6$D$_6$ scintillation detectors. Analyzing the high-resolution capture data by means of the R-matrix code SAMMY, a total of 51 resonances have been identified up to 122 keV. Completing the analysis of the unresolved resonance region by the code SESH, the Maxwellian averaged cross sections have been calculated for the temperature range $kT$ = 5--100 keV of astrophysical importance. Our data confirm the most recent findings by Guber \emph{et al.} \cite{bibN} within $kT$ = 10 -- 60 keV range. The decrease of the new MACS causes an increase of 60\% of the final $s$-process abundance of $^{58}$Ni. However, this change does not propagate to heavier $s$-process isotopes, since $^{58}$Ni is in all cases efficiently destroyed by neutron capture.

\section*{ACKNOWLEDGMENTS}

M. Pignatari acknowledges significant support from NSF grants PHY 02-16783 and PHY 09-22648 (Joint Institute for Nuclear Astrophysics, JINA) and EU MIRG-CT-2006-046520. The continued work on codes and in disseminating data is made possible through funding from the Ambizione grant of the SNSF (MP, Switzerland). M. Pignatari also thanks for support from EuroGENESIS. NuGrid data is served by Canfar/CADC.

T. Rauscher acknowledges the support from the EUROCORES Programme EuroGENESIS, as well as the Swiss NSF. In addition, the support from the ENSAR/THEXO European FP7 Programme and the Hungarian Academy of Sciences are gratefully recognized.

\end{document}